\documentclass[amsmath,superscriptaddress,showpacs,prl,twocolumn,letterpaper] {revtex4}
\usepackage{bm}
\usepackage{bbm}
\usepackage{enumerate}
\usepackage{graphicx}
\usepackage[dvips]{epsfig}
\usepackage{epsf}
\usepackage{physics}
\usepackage{xcolor}
\usepackage[normalem]{ulem}

\makeatletter
\newcommand{\Rmnum}[1]{\expandafter\@slowromancap\romannumeral #1@}
\makeatletter

\begin{document}
\title{Valley current in graphene through electron-phonon interaction}

\author{Ankang Liu}
\email[]{liuankang@tamu.edu}
\affiliation{Department of Physics and Astronomy, Texas A\&M University, College Station, Texas 77843-4242, USA}
\affiliation{Department of Condensed Matter Physics, The Weizmann Institute of Science, Rehovot 76100, Israel}

\author{Alexander M. Finkel'stein}
\affiliation{Department of Physics and Astronomy, Texas A\&M University, College Station, Texas 77843-4242, USA}
\affiliation{Department of Condensed Matter Physics, The Weizmann Institute of Science, Rehovot 76100, Israel}

\begin{abstract}
We discuss valley current, which is carried by quasiparticles in graphene. We show that the valley current arises owing to a peculiar term in the electron-phonon collision integral that mixes the scalar and vector gauge-field-like vertices in the electron-phonon interaction. This mixing makes collisions of phonons with electrons sensitive to their chirality, which is opposite in two valleys. As a result of collisions with phonons, electrons of the different valleys deviate in opposite directions. Breaking the spatial inversion symmetry is not needed for a valley-dependent deviation of the quasiparticle current. The effect exists both in pristine graphene or bilayer graphene samples, and it increases with temperature owing to a higher rate of collisions with phonons at higher temperatures. The valley current carried by quasiparticles could be detected by measuring the electric current using a nonlocal transformer of a suitable design.
\end{abstract}


\maketitle

\emph{Introduction.}\hspace{2mm}Graphene \cite{neto2009electronic,sarma2011electronic} is a two-dimensional (2D) sheet of carbon atoms with a honeycomb lattice. One characteristic of the honeycomb lattice is its band structure which, in the case of pristine graphene, has Dirac cones located at the corners of the first Brillouin zone \cite{wallace1947band}. The Dirac cones at two nonequivalent points of the corners are called the $K$ and $K'$ valleys, respectively. Recently, physicists are more and more interested in the valley-related physics, which forms a new subject called valleytronics \cite{schaibley2016valleytronics}. The control of the valley degrees of freedom could be potentially used for quantum computations and communications.

Systems with honeycomb lattices possess a nonzero Berry curvature, opposite in the two valleys, if the band gap is opened when spatial inversion symmetry is broken \cite{xiao2007valley,xiao2010berry,lensky2015topological}. The nonzero Berry curvature may reveal itself via the valley Hall effect which is reminiscent of the spin Hall effect \cite{kane2005quantum}. Remarkably, some experimental groups have already confirmed that this valley dependent effect could be measured through a nonlocal transport in graphene superlattices \cite{gorbachev2014detecting} or in a dual-gated bilayer graphene sample \cite{shimazaki2015generation,yamamoto2015valley}. Because of an extremely low intervalley scattering rate, the valley current could be detected at distances exceeding 1 $\mu m$.

Still, transport studies which relied on the Berry curvature physics \cite{xiao2010berry} needed a system with broken inversion symmetry and low temperatures. By contrast, in this Rapid Communication we discuss the possibility of working
with a valley current transported by quasiparticles at high temperatures in pristine graphene, both single and double layered. For this purpose, we identified the valley-dependent process in the electron-phonon (\emph{el-ph}) scattering, using the fact that one of the amplitudes of the \emph{el-ph} interaction is sensitive to the chirality of the quasiparticles, which is opposite in the two valleys. By solving the quantum kinetic equation for the \emph{el-ph} scattering in the presence of an external electric field, we demonstrate that the distribution of the quasiparticles contains a term with a quadruple angular dependence, different for the two valleys. In short, current carriers passing through a population of phonons are turned by them in different directions for the two valleys. This opens an opportunity for controlling the valley currents using samples with a designed geometry. The whole effect is owing to transitions between the two sublattices of the honeycomb lattice caused by \emph{el-ph} scattering. The discussed mechanism holds for any honeycomb lattice system.

\emph{Quantum kinetic equation.}\hspace{2mm}In order to study the transport, we derive the quantum kinetic equation in the case of a single-layer graphene. (For bilayer graphene, see Ref. [\onlinecite{SM}].)\nocite{abrikosov2012methods,schutt2011coulomb,kechedzhi2008quantum,mccann2013electronic,ochoa2011temperature} The free-electron and the \emph{el-ph} interaction terms in the Hamiltonian are $H_{e}=\sum_{\bm{p}}\Psi_{\bm{p}}^{\dagger}H_{\bm{p}}\Psi_{\bm{p}}$ and $H_{e-ph}=\sum_{\bm{p},\bm{q}}\Psi_{\bm{p}+\bm{q}}^{\dagger}M_{\bm{q}}\Psi_{\bm{p}}A_{\bm{q}}$, respectively. Here, the fermion operator $\Psi_{\bm{p}}$ is a spinor defined in the sublattice space, and the bosonic field $A_{\bm{q}}=b_{\bm{q}}+b_{-\bm{q}}^{\dagger}$ describes the annihilation and creation of the longitudinal phonons. In this research, we consider the interaction with the longitudinal phonons, because only they provide electrons with a valley-dependent dynamics, which we are interested in.

We first concentrate on one of the valleys. The kinetic term for electrons in the $K$ valley, $H_{\bm{p}}^{K}=v_{F}\bm{\sigma} \cdot \bm{p}-\epsilon_{F}\mathbbm{1}_{2\times2}$, is obtained from the standard tight-binding model for the honeycomb lattice \cite{wallace1947band,neto2009electronic,sarma2011electronic}. Here, $\epsilon_{F}$ is the Fermi energy, and components of the matrix vector $\bm{\sigma}=(\sigma_{x},\sigma_{y})$ are the standard Pauli matrices. We assume that the system is not too close to the neutral point, but at the level of the current carrier concentrations typical for metallic graphene. For the longitudinal acoustic phonons, the matrix elements of the \emph{el-ph} interaction are described by a $2\times 2$ matrix \cite{woods2000electron,suzuura2002phonons,manes2007symmetry},
\begin{align}\label{vertexKPRL}
M_{\bm{q}}^K=\vert\bm{q}\vert\sqrt{\frac{1}{2\rho\omega_{\bm{q}}}}
\begin{pmatrix}
	g_{1}&g_{2}e^{i(2\theta_{\bm{q}}+3\eta)}\\
	g_{2}e^{-i(2\theta_{\bm{q}}+3\eta)}&g_{1}\\
\end{pmatrix}.
\end{align}	
We are mostly interested in the details of the off-diagonal elements of this matrix. Here, $\theta_{\bm{q}}$ is the angle between the phonon's momentum $\bm{q}$ and the $x$ direction, while $\eta$ is the angle of the $x$ direction measured from the zigzag direction of the honeycomb lattice. The combination in front of the parentheses is standard for the \emph{el-ph} interaction with acoustic phonons: $\rho$ is the mass density of the graphene sample, and $\omega_{\bm{q}}=v_{s}|\bm{q}|$ is the phonon frequency for the longitudinal acoustic mode with $v_{s}\ll v_{F}$.

In matrix $M_{\bm{q}}$, the diagonal coupling constant $g_{1}$ comes from the deformation potential (DP), and before screening has a bare value of 20-30 eV \cite{suzuura2002phonons}. The magnitude of $g_{2}$ has been estimated to be 1.5 eV. The term affiliated with $g_{2}$ reveals some similarities with the vector potential (VP) for the electromagnetic field \cite{suzuura2002phonons,manes2007symmetry,mariani2010temperature,chen2012electron}.

\begin{figure}[h] \centerline{\includegraphics[clip, width=1  \columnwidth]{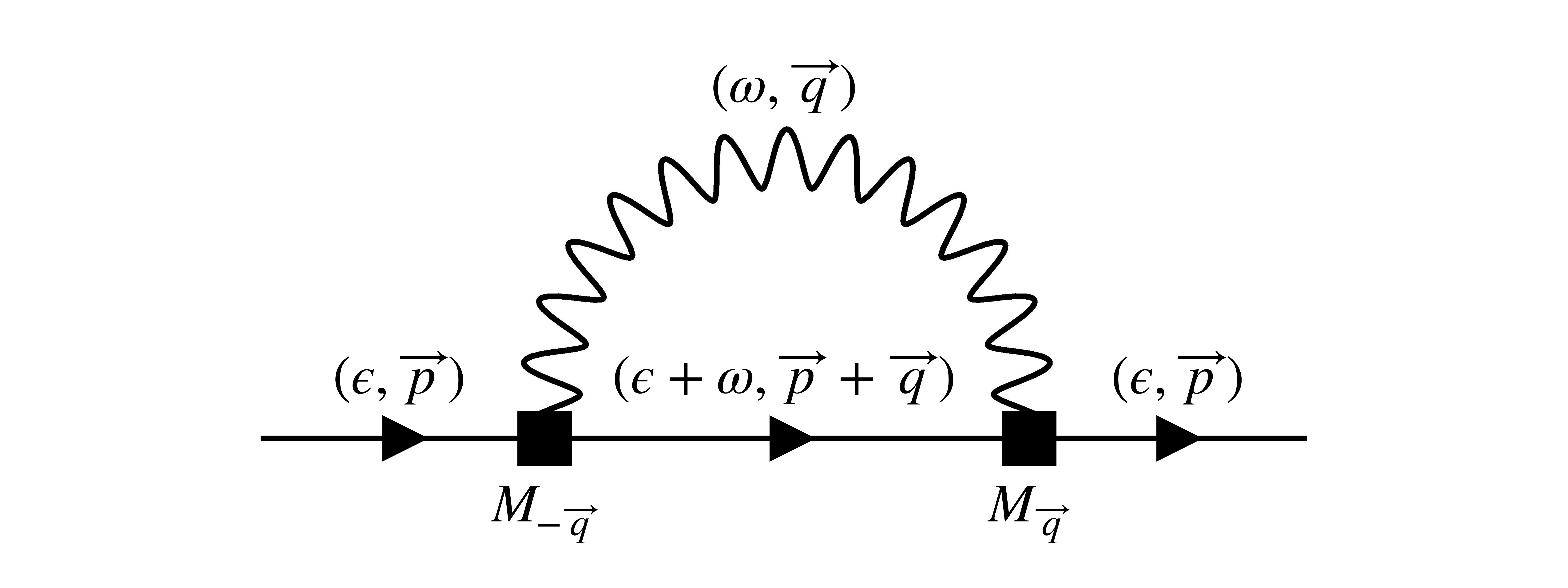}}
	
\protect\caption{The electronic self-energy due to the \emph{el-ph} interaction. The solid line represents the electron propagator while the wavy line means the phonon propagator. The black square here is the full \emph{el-ph} vertex, which contains both the scalar and vector gauge-field-like part of the vertex.}
	
\label{fig:diagram}
	
\end{figure}

To obtain the \emph{el-ph} collision integral in the quantum kinetic equation, we consider the self-energy diagram presented in Fig. \ref{fig:diagram}.
In the derivation of the quantum kinetic equation we apply the quasiclassical approximation \cite{prange1964transport,kamenev2011field,rammer1986quantum}. We rely on the fact that for the \emph{el-ph} interaction the self-energy has a weak dependence on $\xi_{\bm{p}}=v_{F}p-\epsilon_{F}$. In the quasiclassical approximation, the electron momentum $\bm{p}$ is placed on the Fermi surface. This could be achieved by integrating the Green's functions with respect to $\xi_{\bm{p}}$: $g^{K/R/A}=\frac{i}{\pi}\int d\xi_{\bm{p}}G^{K/R/A}$. The remaining dependence on an electron direction is described by a unit vector $\bm{n_{p}}=\bm{p}/p$. The reason to use quasiclassics is that we are interested in the effects related to the angular dependencies.

The \emph{el-ph} collision integral $I_{e-ph}(f,n)$ has to be written in terms of $f$, the quasiclassical
distribution function of electrons, and the distribution function of phonons $n$.
For our purposes it will be enough to assume that phonons are under thermal equilibrium, i.e., $n=n_{0}(\omega)=(e^{\omega/T}-1)^{-1}$. The \emph{el-ph} collision integral for electrons in the $K$ valley contains (among others) the following specific term \cite{SM},
\begin{align}\label{collision_int_1PRL}
	I_{e-ph}^{v}(f,n)&=2\pi\nu_{0}\int d\epsilon'\int\frac{d\theta_{\bm{p}'}}{2\pi}\alpha(q)g_{1}g_{2}(\bm{n_{p}}+\bm{n_{p'}})\cdot\bm{d}
	\nonumber
	\\
    &\times\big[f'(1-f)(1+n)-f(1-f')n\big](\delta_{+}-\delta_{-}).
\end{align}
Here, $\nu_{0}=p_{F}/2\pi v_{F}$ is the density of state for the electron per valley and per spin, $\alpha(q)=q^{2}/2\rho\omega_{\bm{q}}$, and $\bm{d}=(cos2\theta_{\bm{q}},-sin2\theta_{\bm{q}})$ is a unit vector associated with the vector gauge-field-like vertex. Without loss of generality, we take here $\eta=0$ assuming that our $x$ direction is along the so-called zigzag lattice direction. We also use short notations here: $f=f(\epsilon,\bm{n_{p}})$, $f'=f(\epsilon',\bm{n_{p'}})$, $n=n_{0}(\epsilon'-\epsilon)$, and $\bm{q}=p_{F}(\bm{n_{p'}-n_{p}})$. Energy conservation in the collision integral is controlled by $\delta_{\pm}=\delta(\epsilon'-\epsilon\mp v_{s}p_{F}\vert\bm{n_{p'}-n_{p}}\vert)$.

\emph{Transformation from $K$ to $K'$ valley.}\hspace{2mm}For electrons in the $K'$ valley, the kinetic term $H_{\bm{p}}^{K'}=v_{F}(-p_{x}\sigma_{x}+p_{y}\sigma_{y})-\epsilon_{F}\mathbbm{1}_{2\times2}$ \cite{wallace1947band,neto2009electronic,sarma2011electronic}. Next, the \emph{el-ph} interaction is given by the matrix $M_{\bm{q}}^{K'}$, which is connected to the $M_{\bm{q}}^{K}$ through the relation $M_{\bm{q}}^{K'}=\left(M_{-\bm{q}}^{K}\right)^*$ \cite{manes2007symmetry,suzuura2002phonons}. One may observe that the transformation from the $K$ to $K'$ valley could be achieved by using the following substitutions (again, we set here $\eta=0$):

\noindent (i) $\hat{\bm{v}}^{K}=v_{F}(\sigma_{x},\sigma_{y})\rightarrow\hat{\bm{v}}^{K'}=v_{F}(-\sigma_{x},\sigma_{y})$; (ii) $\bm{n_{p}}^{K}=(cos\theta_{\bm{p}},sin\theta_{\bm{p}})\rightarrow\bm{n_{p}}^{K'}=(-cos\theta_{\bm{p}},sin\theta_{\bm{p}})$;
\noindent and (iii) $\bm{d}^{K}=(cos2\theta_{\bm{q}},-sin2\theta_{\bm{q}})\rightarrow\bm{d}^{K'}=(cos2\theta_{\bm{q}},sin2\theta_{\bm{q}})$.

With the use of these transformations, one could check that the collision term in the quantum kinetic equation remains unchanged for the $K'$ valley, except that the $g_{1}g_{2}$ term presented by Eq. (\ref{collision_int_1PRL}) acquires an opposite sign. We ascribe the peculiarity of this valley-dependent term to the origin of the vector gauge-field-like vertex (VP). Indeed, the $g_{2}$ term, as the off-diagonal part of the \emph{el-ph} vertex, comes from the intersublattice hopping mediated by the lattice vibrations. Consequently, this term is sensitive to the direction of the quasimomentum as well as to the chirality of electrons and, hence, is different in two valleys. On the contrary, the scalar $g_{1}$ term is the on-site energy, which is valley independent. Therefore, only the mixture of $g_{1}$ and $g_{2}$ terms would produce a valley-contrasting term $(\bm{n_{p}\cdot\bm{d}})^{K/K'}$ in the collision integral that leads to a valley-dependent dynamics.

\emph{Valley-dependent dynamics under an external electric field.}\hspace{2mm}In the presence of an electric field, $\bm{E}=-\bm{\nabla}\Phi_{\bm{E}}$, we parametrize $\delta f$, a small deviation of the electron distribution function from the local equilibrium $f_{0}=(e^{(\epsilon-e\Phi_{\bm{E}})/T}+1)^{-1}$, as $\delta f=(-\frac{\partial f_{0}}{\partial\epsilon})\varphi$ \cite{landau1981course}. In the linear response regime, we have to solve the equation for the steady solution $\varphi$,
\begin{align}\label{single_1PRL}
v_{F}e\bm{E}\cdot\bm{n_{p}}\frac{\partial f_{0}}{\partial\epsilon}=I_{0}(\varphi)+[I_{e-ph}^{v}]^{K/K'}(\varphi).
\end{align}
Here $I_{0}(\varphi)=\frac{\partial f_{0}}{\partial\epsilon}\varphi/\tau$ is a valley-independent collision term written in the relaxation time approximation. $I_{0}(\varphi)$ has been introduced to account for the valley-independent scatterings which determine the conventional transport properties of the system, e.g., the electron-impurity scattering. For simplicity, we will assume here that the valley-independent scattering is mostly of a short-ranged character. i.e., the relaxation time $\tau$ is same for harmonics of different orders \cite{knap1996weak}. The other term in Eq. (\ref{single_1PRL}) describes the valley-dependent part of the collision integrals [\emph{cf}. Eq. (\ref{collision_int_1PRL})],
\begin{align}\label{single_3PRL}
[I_{e-ph}^{v}]^{K/K'}(\varphi)&=\pm2\pi\nu_{0}\int d\epsilon'\int\frac{d\theta_{\bm{p}'}}{2\pi}\alpha(q)
\nonumber
\\
&\times[g_{1}g_{2}(\bm{n_{p}}+\bm{n_{p'}})\cdot\bm{d}](\frac{\partial n_{0}}{\partial\omega})\Big|_{\omega=\epsilon'-\epsilon}
\nonumber
\\
&\times(f_{0}'-f_{0})(\varphi'-\varphi)(\delta_{+}-\delta_{-}).
\end{align}
Here, $\pm$ refers to the $K/K'$ valleys.

One can show that Eq. (\ref{single_1PRL}) can be solved perturbatively \cite{SM}, assuming $\varphi$ to be $\varphi\simeq\varphi_{0}+\varphi_{1}^{K/K'}$ with $\varphi_{0}\gg\varphi_{1}^{K/K'}$. Then, the functions $\varphi_{0}$ and $\varphi_{1}^{K/K'}$ have to satisfy two iteration equations,
\begin{align}\label{single_2PRL}
&v_{F}eE\frac{\partial f_{0}}{\partial\epsilon}cos\theta_{\bm{p}}=I_{0}(\varphi_{0}),
\nonumber
\\
&I_{0}(\varphi_{1}^{K/K'})+[I_{e-ph}^{v}]^{K/K'}(\varphi_{0})=0.
\end{align}
We seek for a Drude-kind solution $\varphi_{0}=A_Ecos\theta_{\bm{p}}$ with $A_E=eEl$, where the mean free path length $l\equiv v_{F}\tau$. Then, the term mixing the two vertices in the \emph{el-ph} interaction generates a nontrivial angular dependence in the distribution function,
\begin{align}\label{single_solution_anglePRL}
\varphi_{1}^{K/K'}=\mp[B_{ph}^{(2)}cos(2\theta_{\bm{p}})+B_{ph}^{(4)}cos(4\theta_{\bm{p}})]
\end{align}
with $B_{ph}^{(2/4)}=(\Gamma_{e-ph}^{(2/4)}\tau) A_E$. [Here, and in Eq. (\ref{single_2PRL}), we took for simplicity $\bm{E}=E(1,0)$ to be along the zigzag direction for which $\eta=0$.] It turned out that the fourth harmonic (for numerical reasons) yields only a negligible correction to the effect that we are interested in. We therefore omit the fourth harmonic in the consideration below. The appearance of the quadruple valley-dependent term in the distribution function of the current-carrying state is the central observation of this Rapid Communication.

To utilize the valley-dependent angular distribution generated by the \emph{el-ph} scattering,
we will be interested in relatively high temperatures, e.g., room temperature and above, when $T\gg T_{BG}$; $T_{BG}\equiv2v_{s}p_{F}$ is the Bloch-Gr\"uneisen temperature. (For the sake of convenience, we introduce now a dimensionless concentration $\tilde{n}$, so that the electron concentration $n=\tilde{n}\times10^{12}\ cm^{-2}$. Then, the Bloch-Gr\"uneisen temperature depends on $\tilde{n}$ as $T_{BG}\approx57\times\sqrt{\tilde n}\ K$.) For temperatures much exceeding $T_{BG}$, the rate $\Gamma_{e-ph}^{(2)}$ is estimated to be \cite{SM}
\begin{align}\label{gammaelph2PRL}
\Gamma_{e-ph}^{(2)}\simeq3\times10^{-3}\tilde{n}\left(\frac{T}{T_{BG}(\tilde{n})}\right)\ {ps}^{-1}.
\end{align}

\begin{figure}[h] \centerline{\includegraphics[clip, width=1  \columnwidth]{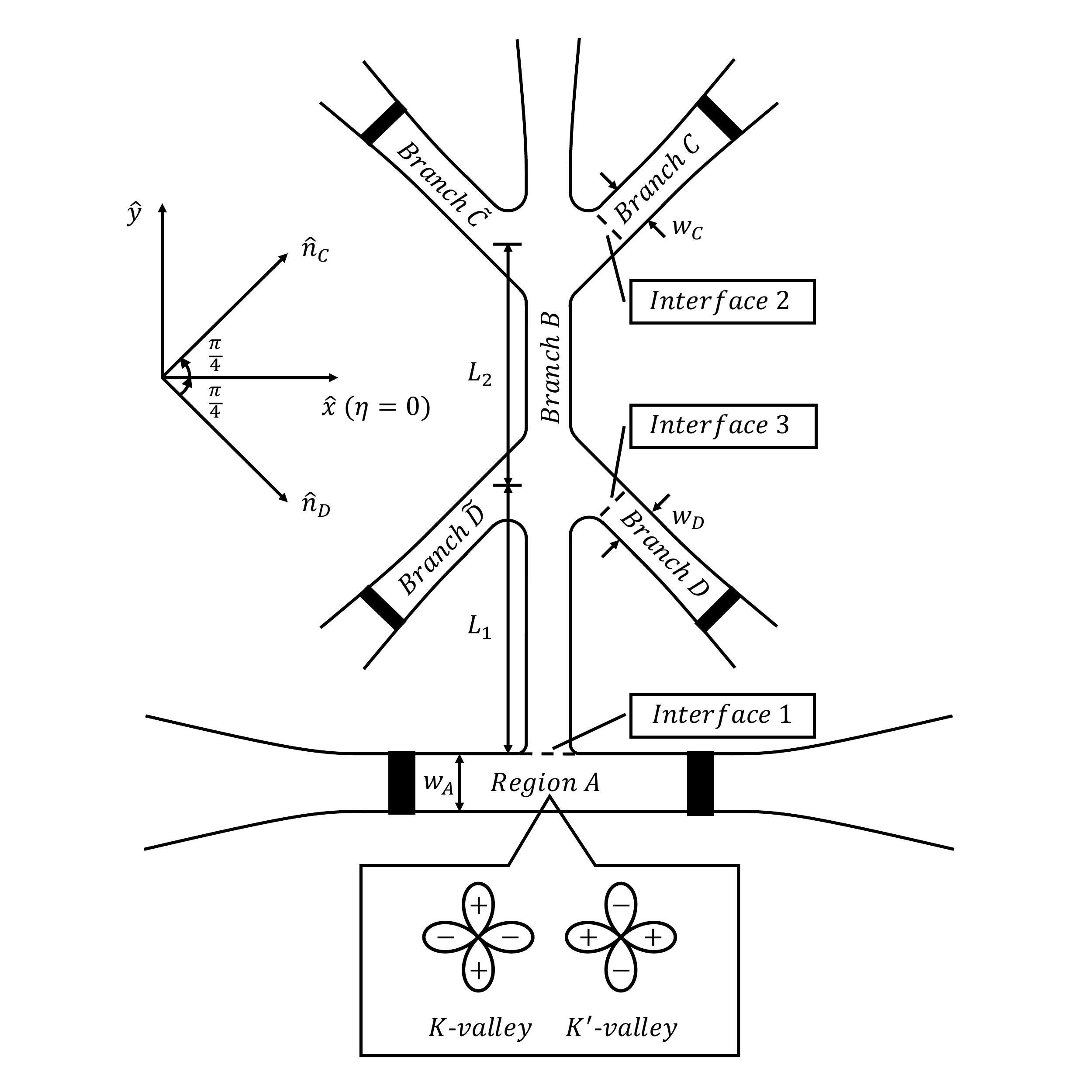}}
	
	\protect\caption{The geometry of a graphene sample suggested for the generation and detection of the valley current. At the bottom, the valley-dependent quadruple distributions caused by the \emph{el-ph} scattering inside the region $\cal A$ are shown, where the $\pm$ sign means the excess and deficit of the distributed carriers with certain momentum directions.}
	
	\label{fig:geometry}
	
\end{figure}

\emph{Generation and detection of the valley current.}\hspace{2mm}We are ready to show how the valley current arises as a result of the valley-dependent $\varphi_{1}^{K/K'}$, and suggest a scheme of detecting a nonlocal signal. Let us consider the geometry presented in Fig. \ref{fig:geometry}.
In region $\cal A$, the electric current flows along the $x$ direction. As we have shown, the distribution function contains the valley-dependent quadruple term $\varphi_{1}^{K/K'}$. In a sample with the discussed geometry, $\varphi_{1}^{K/K'}$ leads to a valley current that propagates along branch $\cal B$. Indeed, let us consider interface 1 (a conditional boundary between regions $\cal A$ and $\cal B$). The carriers with $\pi\ge\theta_{\bm{p}}\ge0$ leave region $\cal A$, pass through the interface, and enter region $\cal B$. Consequently, the distribution $\varphi_{1}^{K}$ provides a nonzero upward flux through interface 1 and, eventually, an upward flow of the $K$-valley carriers. The carriers in the $K'$ valley would react oppositely. Finally, there will be a valley current along branch $\cal B$. This is similar to the injection of the spin current in spintronic devices.

The distribution functions that introduce the fluxes of the $K$- and $K'$-valley carriers inside branch $\cal B$, after a few scattering events, acquire the Drude-like form directed oppositely for two valleys. In other words, at a distance $l_{eq}\agt l$, but much less than the intervalley scattering length $l_{v}$, one can introduce the valley-dependent chemical potentials $\mu^{K/K'}=\bar{\mu}+\delta\mu^{K/K'}(y)$, and apply the diffusion approximation for the flux-carrying particles. Note that because of the electroneutrality, $\delta\mu^{K}=-\delta\mu^{K'}$.
In order to maintain a stationary valley current flow deep inside the branch $\cal B$, $\delta\mu^{K/K'}$ would have a linear spatial dependence, i.e., $\bm{\nabla}(\delta\mu^{K/K'})=\mp l^{-1}C_{\cal B}\bm{\hat{y}}$. The distribution functions $\varphi^{K/K'}_{\cal B}$ corresponding to the valley-current state are $\varphi^{K/K'}_{\cal B}=\pm C_{\cal B}sin\theta_{\bm{p}}$. Finally, this gives the valley current density $\bm{j}_{v}\equiv\bm{j}^{K}-\bm{j}^{K'}=4\sigma_{0}(el)^{-1}C_{\cal B}\bm{\hat{y}}$ along branch $\cal B$ with $\sigma_{0}=\frac{1}{2}e^2\nu_{0}v_{F}^{2}\tau$ to be the Drude conductivity for graphene per valley and per spin.

It remains to get an estimate for $C_{\cal B}$. For that we have to match the valley current on the $\cal A$ side of the interface line  with that on the $\cal B$ side. To analyze the question in full detail, one has to solve the so-called diffuse emission problem for a given geometry (see, e.g., Refs. [\onlinecite{morse1954methods,shekhter2005diffuse}]). We, however, limit ourselves to a qualitative discussion only. For a qualitative estimate, we use the distribution functions $\varphi_{1}^{K/K'}$ below the interface line, while above the line we take the distribution functions $\varphi^{K/K'}_{\cal B}$, i.e.,
\begin{align}\label{BCPRL}
\int_{0}^{\pi}sin\theta_{\bm{p}}\varphi_{1}^{K/K'}d\theta_{\bm{p}}
\simeq \int_{0}^{2\pi}sin\theta_{\bm{p}}\varphi^{K/K'}_{\cal B}d\theta_{\bm{p}}.
\end{align}
Eventually, we get $C_{\cal B}\simeq2B_{ph}^{(2)}/3\pi$ as an estimate for $C_{\cal B}$.

Now, let us discuss the mechanism of detecting the valley current. The main point here is that the valley current carriers inside the branch $\cal B$, in the process of collisions with phonons, generate a new term $\tilde{\varphi}_{\cal B}$ in the electron distribution function, with a nontrivial angular dependence. By solving the kinetic equation, we obtain (see Ref. [\onlinecite{SM}] for details)
\begin{align}\label{valley_independent_distribution_in_B}
\tilde{\varphi}_{\cal B}=D_{ph}^{(2)}sin(2\theta_{\bm{p}})
\end{align}
with $D_{ph}^{(2)}=(\Gamma_{e-ph}^{(2)}\tau)C_{\cal B}$. Because of the angular dependence of the distribution $\tilde{\varphi}_{\cal B}$, we expect to get a current flux injected into the side-directed branches chosen for recording.

In the discussed geometry, we suggest to measure electric currents flowing in the opposite directions (from left to right and from right to left) in two pairs of side branches $(\cal C,\tilde{\cal C})$ and $(\cal D,\tilde{\cal D})$, as shown in Fig. \ref{fig:geometry}. Following the above discussion, the distribution function inside each of the branches, after a few collisions, acquires the Drude form. For example, inside branch $\cal C$ the function $\varphi_{\cal C}$ acquires the form $\varphi_{\cal C}=elE_{\cal C}cos(\theta_{\bm{p}}-\frac{\pi}{4})$. The combination $elE_{\cal C}$ could be estimated by matching the fluxes on both sides of interface 2. Similarly to Eq. (\ref{BCPRL}), we get $elE_{\cal C}\simeq2D_{ph}^{(2)}/3\pi$. Next, for branch $\cal D$, the injection yields the opposite sign, i.e., $E_{\cal D}=-E_{\cal C}$. Utilizing the chain of relations which connect $A_{E}$ with $B_{ph}^{(2)}$, $C_{\cal B}$, $D_{ph}^{(2)}$, and, finally, with $E_{\cal C}$ and $E_{\cal D}$, we can estimate the current density ratio for our design of the nonlocal transformer,
\begin{align}\label{non-local_resistancePRL}
\frac{j_{\cal C}}{j_{\cal A}}=-\frac{j_{\cal D}}{j_{\cal A}}\simeq\frac{E_{\cal C}}{E}\simeq\frac{4}{9\pi^{2}}\left(\Gamma_{e-ph}^{(2)}\tau\right)^{2}.
\end{align}

\emph{Discussion and conclusion.}\hspace{2mm}In the remaining part of this Rapid Communication, we estimate the typical value of $j_{\cal C}/j_{\cal A}$. To do this, we extract the scattering time $\tau$ from the conductivity $\sigma_0$, by simply using the Drude formula, i.e., $\tau=\sigma_{0}/\frac{1}{2}e^{2}\nu_{0}v_{F}^{2}$. Consequently, we find
\begin{align}\label{nonlocalCPRL}
	\frac{j_{\cal C}}{j_{\cal A}}\simeq0.6\times10^{-10}\left(\frac{\sigma_{0}(n)}{\sigma_{q}}\right)^{2}
	\left(\frac{T}{T_{BG}(\tilde{n}=1)}\right)^{2}.
\end{align}
Here, the quantum conductivity $\sigma_{q}\equiv e^{2}/h$ is introduced as the unit of the conductivity. For metallic samples with a usual conductivity $\sigma_{0}(n)$ \cite{tan2007measurement,sarma2011electronic}, the current density ratio could be expected in the region $10^{-7}-10^{-6}$. In the presence of a mismatch of the sample orientation, i.e., when $\eta\neq0$, the discussed nonlocal effect survives. It is suppressed only by a geometric factor $cos^{2}(3\eta)$ \cite{SM}.

Our consideration was limited to the case of degenerate electrons and the assumption that the intervalley scattering is negligible. Both assumptions limit the temperature from above. Still, because of a large difference between $v_s$ and $v_F$ in graphene, there remains a substantial interval of temperatures, $T_{BG}\ll T\ll\epsilon_{F}$, that could be addressed provided that $\tilde{n}$ is not so small. 

Owing to the fact that the valley scattering length $l_{v}\gtrsim1\ \mu m$, we expect that the current density ratios $j_{\cal C}/j_{\cal A}$ and $j_{\cal D}/j_{\cal A}$ could be measured through the geometry suggested by us in Fig. \ref{fig:geometry} with $1\ \mu m\gtrsim L_{1}, L_{2}\gg l\simeq10\ nm$. The strong inequality here is needed in order to reliably prevent the penetration of particles from region $\cal A$ straight into branches $(\cal C,\tilde{\cal C})$ and $(\cal D,\tilde{\cal D})$.

To conclude, we have argued that a valley current carried by quasiparticles could be generated and detected through a properly arranged geometric design. Our scheme relies on the fact that the term in the \emph{el-ph} collision integral originating from a mixture of the scalar and vector gauge-field-like vertices has an opposite sign for the $K$ and $K'$ valleys. The effectiveness of the discussed mechanism grows with temperature by virtue of a greater \emph{el-ph} collision rate at higher temperatures. In these respects, it differs entirely from the Berry curvature mechanism which works when the system is not far from the ground state, which has to be insulating. In short, our study provides an alternative approach to generate and detect the long-range propagating valley current in a pristine single- or double-layered graphene sample, which does not require the breaking of spatial inversion symmetry.

According to the estimates presented in this work, a current density ratio of the designed nonlocal transformer is small, but could be detected. The point is that the discussed mechanism does not have fragile elements, and is not sensitive to noise. Furthermore, since the quadruple character of the distribution is very specific, it can be checked by measuring currents in different branches. For example, if the information about the width of the branches, $w_{\cal C}$ and $w_{\cal D}$, is available, one may expect that $[(I_{\cal C}/w_{\cal C})-(I_{\cal D}/w_{\cal D})]\gg[(I_{\cal C}/w_{\cal C})+(I_{\cal D}/w_{\cal D})]$, and so on. In this way, one could exclude a parasitic signal that may come from a leakage from the region $\cal A$ to one of the branches. In this Rapid Communication, the generation and detection of the valley current was discussed in terms of the transport measurements. Alternatively, one can try to detect the valley polarization $\mu^{K}-\mu^{K'}$ which arises as a consequence of the injection of valley current inside branch $\cal B$ without introducing the side branches. Indeed, the valley polarization $\mu^{K}-\mu^{K'}$ in our scheme reaches the range of a few $meV$ at a distance of order $l_v$ from interface 1. Polarization of this scale can be detected in a pool at the end of branch $\cal B$ by the method considered in Ref. [\onlinecite{wehling2015probing}] or by magneto-oscillations (see, e.g., Ref. [\onlinecite{abrikosov2017fundamentals}]).

From a fundamental point of view, the discussed mechanism, which holds generally for any honeycomb lattice system, demonstrates that the valley current can be of a kinetic origin, rather than be obligatorily related with the Berry curvature physics. Next, it opens a perspective to study the nontrivial aspects of the \emph{el-ph} interaction, and the intervalley scattering rate at high temperatures. We expect this research could open up another unexplored possibility in the area of valleytronics.\\

\begin{acknowledgments}
\emph{Acknowledgments.}\hspace{2mm}We thank I. Borzenets, J. Sinova, K. Tikhonov, and V. Zyuzin for their interest in this work. We gratefully acknowledge I. Gornyi for helpful information. A. L. thanks the Department of the Condensed Matter Physics at the Weizmann Institute of Science for hospitality. The work was supported by research grants from the Veronika A. Rabl Physics Discretionary Fund and the Benoziyo Endowment Fund for the Advancement of Science.
\end{acknowledgments}

\bibliographystyle{apsrev}
\bibliography{MyBIB}

\end{document}